\documentclass[aps,amssymb,amsmath,twocolumn,secnumarabic]{revtex4}

\usepackage[dvips]{graphicx}
\usepackage{amsfonts}
\usepackage{amsmath}
\usepackage{mathrsfs}
\usepackage{xypic}

\usepackage[arrow, matrix, curve]{xy}
\xymatrixcolsep{0.5cm} \xymatrixrowsep{0.5cm}

\def\L2{{\it Fun\/}\bigl(\text{\GL}\bigr)}

\newcommand{\gl}{\ensuremath{\text{gl}(1|1)}}
\newcommand{\HH}{H}
\newcommand{\G}{G}

\newcommand{\cm}{c_-}
\newcommand{\cp}{c_+}

\newcommand{\del}{\partial}

\newcommand{\Ad}{\text{Ad}}

\def\gl{{\rm gl(1$|$1)}}
\def\GL{{\rm GL(1$|$1)}}

\newcommand{\g}{\mathfrak{g}}
\newcommand{\aplus}{\mathfrak{a_+}}
\newcommand{\am}{\mathfrak{a_-}}
\newcommand{\apm}{\mathfrak{a_\pm}}
\newcommand{\azero}{\mathfrak{a_0}}

\newcommand{\h}{\mathfrak{h}}

\newcommand{\str}{\text{str}}

\newcommand{\nn}{\mathfrak{n}}
\renewcommand{\k}{\mathfrak{k}}
\newcommand{\s}{\mathfrak{s}}

\def\GL{{\rm GL(1$|$1)}}

\newcommand{\beqa}{\begin{eqnarray}}
\newcommand{\eeqa}{\end{eqnarray}}

\DeclareMathOperator{\sdim}{sdim}

\begin{document}

\title{$\mathcal{N}=2$ Superconformal Symmetry in Super Coset Models}

\author{Thomas Creutzig, Peter B. R\o nne, Volker Schomerus}

\affiliation{
DESY Theory Group, DESY Hamburg, Notkestrasse 85, D-22603 Hamburg,
Germany}

\email{thomas.creutzig@desy.de, peter.roenne@desy.de, volker.schomerus@desy.de}

\date{April 2009}

\begin{abstract}
We extend the Kazama-Suzuki construction of models with $\mathcal{N}=(2,2)$
world-sheet supersymmetry to cosets $S/K$ of supergroups. Among
the admissible target spaces that allow for an extension to $\mathcal{N}=2$
superconformal algebras are some simple Lie supergroups, including
PSL(N$|$N). Our general analysis is illustrated at the example of
the $\mathcal{N}=1$ WZNW model on GL(1$|$1). After constructing its $\mathcal{N}=2$
superconformal algebra we determine the (anti-)chiral ring of the
theory. It exhibits an interesting interplay between world-sheet
and target space supersymmetry.
\end{abstract}

\maketitle

\vspace*{-7.0cm} \noindent
{\tt DESY 09-111}\\[6cm]
%{\tt DESY 09-111  \hspace*{12.7cm} arXiveyymm.nnnn}\\[6cm]

\section{Introduction}
Sigma models with target superspaces have appeared in
a large variety of physics problems, ranging from
$\mathcal{N}=4$ super Yang-Mills theory to disordered electron systems.
In this note we are particularly interested in theories
for which an explicit $\mathcal{N}=1$ superconformal symmetry
on the world-sheet gets enhanced to $\mathcal{N}=2$. A few
basic examples have been discussed in the literature.
These include the supersymmetric sigma model on the
so-called twistorial Calabi-Yau $\mathbb{C}P^{3|4}$
that featured in Witten's work \cite{Witten:2003nn}
on twistor string theory (see e.g.\ \cite{Aganagic:2004yh,
Kumar:2004dj,Ahn:2004xs,Belhaj:2004ts,Ricci:2005cp,Seki:2006cj}).
Sigma models on Calabi-Yau superspaces were also
conjectured to describe the mirror partner of string
theory on rigid Calabi-Yau manifolds \cite{Sethi:1994ch,
Schwarz:1995ak}. This makes it seem worthwhile to look for more
general constructions of such models.

Quantum field theories with $\mathcal{N}=2$ superconformal symmetry
possess an intimate and well known relation with topological
field theories. In $\mathcal{N}=(2,2)$ superconformal models, the chiral
Virasoro field $T$ is part of a multiplet involving two
fermionic fields $G^\pm$ with conformal weight $h_G = 3/2$
and a bosonic U(1) current $U$ with relations
\begin{eqnarray*}
  G^+(z) G^-(w)&\sim&\frac{c/3}{(z-w)^3} +
  \frac{U(w)}{(z-w)^2} + \frac{(T+\frac{1}{2}\del U)(w)}{(z-w)}
 \\[2mm]
U(z) \, G^\pm(w) & \sim &  \frac{\pm G^\pm(w)}{(z-w)}\quad ,\ \
U(z)\, U(w) \, \sim \,  \frac{c/3}{(z-w)^2}\, .
\end{eqnarray*}
The same algebra is satisfied by the anti-chiral partners $\bar T,
\bar G^\pm$ and $\bar U$. Given this structure, one may go through
a process of twisting. It results in two different topological
conformal field theories that are known as the A- and B-model,
respectively.

In \cite{Kazama:1988qp} (see also \cite{Spindel:1988sr}
for earlier related work), Kazama and Suzuki described a simple
construction providing many key examples of world-sheet theories
with $\mathcal{N}=2$ superconformal symmetry. They started from an
$\mathcal{N}=1$ Wess-Zumino-Novikov-Witten (WZNW) model for the coset space
$S/K$ and investigated under which conditions the $\mathcal{N}=1$ symmetry
could be extended to an $\mathcal{N}=2$ superconformal algebra. Within the
list of cases they worked out are the $\mathcal{N}=2$ minimal models. These
feature as building blocks for  Gepner's construction of string
theory on Calabi-Yau manifolds.
Our aim here is to generalize the analysis of Kazama and Suzuki to the
case of coset superspaces $S/K$ where both $S$ and $K$ can be
Lie supergroups. Following \cite{Parkhomenko:1992dq,Getzler:1993py},
we shall describe the $\mathcal{N}=2$ superconformal algebras in terms of
supersymmetric Manin triples. Among the resulting
$\mathcal{N}=(2,2)$ theories, we find one family of particular interest: It is
shown that the $\mathcal{N}=1$ WZNW models on the simple supergroups
$S =$ PSL(N$|$N) (with trivial denominator $K = \{ e\}$)
possess an $\mathcal{N}=2$ superconformal symmetry. A related observation
for an $\mathcal{N}=(1,1)$ model on the bosonic base of PSL(2$|$2) was
made and studied by several authors
\cite{Rastelli:2005ph,Dabholkar:2007ey,Gaberdiel:2007vu}.

Let us briefly describe the content of this note. In the next
section we shall outline the construction of $K$-gauged $\mathcal{N}=1$ WZNW
models on a supergroup $S$. As in the case of bosonic targets, the
$S/K$ coset model can be realized within the WZNW model on the
product supergroup $G = S \times K$. We continue by introducing the
notion of a Manin triple for supergroups $G$ and provide a few
examples of this algebraic structure. From the data of a Manin
triple we shall construct the fields $G^\pm$ and $U$ of the $\mathcal{N}=2$
superconformal algebra in section 4. There we also discuss
possible deformations of the $\mathcal{N}=2$ superconformal algebra. In
section 5 we consider $S =$ GL(1$|$1) and $K = \{ e\}$ as a
simple example in which we can easily determine the chiral
ring. The latter is shown to consist of fields in atypical
multiplets of the target space supersymmetry gl(1$|$1). Finally,
we discuss a few extensions and open problems.

\section{Gauged $\mathcal{N}=1$ WZNW models}

WZNW models on coset superspaces with
$\mathcal{N}=1$ world-sheet supersymmetry possess a manifestly supersymmetric
formulation in terms of superfields of the form
\begin{equation}
    \G\ = \ \exp (i\theta\chi)\ g \ \exp (-i\bar\theta\bar\chi)\, .
\end{equation}
Here, $g = g(z,\bar z)$ is a field that takes
values in the supergroup $S$ and $\chi = \chi^a t_a$ is a Lie
super-algebra valued field. The components $\chi^a$ are
fermionic for even generators $t_a$, i.e.\ when $|a| =0$, and they
are bosonic otherwise. The multiplets $\chi^a$ and $\bar
\chi^a$ each transform in the adjoint of the Lie superalgebra $\s$
of $S$.
One may now use the superfield $\G$ along with the covariant
derivatives on the world-sheet given by
\begin{equation}
    D \ = \  -i\frac{\del}{\del \theta} - 2 \theta \del\qquad\text{and}\qquad
    \bar D \ = \  -i\frac{\del}{\del \bar \theta} - 2 \bar \theta \bar
    \del
\end{equation}
to build the usual action of the WZNW model on the supergroup.
Writing down the action also requires fixing some non-degenerate
invariant bilinear form $(\cdot,\cdot)$ on the Lie superalgebra
$\s$. When written in components, the action becomes
\begin{eqnarray}\label{eq:N=1action}
        S^{\mathcal{N}=1}_{\rm WZNW}[\G]& = & S'_{\rm WZNW}[g]\ +\\[2mm]
        & & + \  \frac{1}{2\pi}\int d^2z\, (\chi,\bar\del\chi)+
        (\bar\chi,\del\bar\chi) \ .  \nonumber
\end{eqnarray}
In our notation, the level $k$ of the model is absorbed into the
definition of the bi-linear form $(\cdot,\cdot)$. The formula for
the WZNW action on the $S$-valued field $g$ has the usual form, but
with the bi-linear form $(\cdot , \cdot )$ shifted by half the Killing
form $\langle \cdot , \cdot \rangle$, i.e.\ $(\cdot, \cdot)' =
(\cdot,\cdot) + \frac{1}{2}\langle \cdot,\cdot\rangle$. The Killing form is
constructed and normalized in the standard fashion. In case the
Killing form is proportional to $(\cdot, \cdot)$, the shift of the
bilinear form simply amounts to shifting the level by the dual
Coxeter number. The global target supersymmetry of the $\mathcal{N}=(1,1)$
theory gives rise to holomorphic currents $J^a$ and $\bar J^a$
which satisfy the usual super-symmetric current algebra at level
$k$. These currents include terms that are constructed out of the
fields $\chi^a$ and $\bar \chi^a$. For a simple Lie superalgebra
$\s$, the total central charge of the model is
$$ c \ =\  \frac{(k-h^\vee_\s) \sdim \s}{k} + \frac12
\sdim\s \ = \ \left(\frac32 - \frac{h^\vee_\s}{k}\right) \, \sdim
\s \ .
$$
The second term is the contribution from the fields $\chi^a$. Note
that all these fields possess conformal weight $h_\chi = 1/2$ so
that each fermionic component of $\chi$ contributes $\delta c =
1/2$ to the central charge while each bosonic component subtracts
the same amount.

The gauged WZNW model of Lie groups has been described in e.g.
\cite{Faddeev:1985iz, Bardakci:1987ee, Gawedzki:1988hq, Gawedzki:1988nj, Karabali:1989dk, Karabali:1988au, Tseytlin:1993my}.
The formulation extends immediately to Lie
supergroups. Let $A = A(z,\bar z ,\theta,\bar\theta)$ and $\bar A
= \bar A(z,\bar z,\theta,\bar\theta)$ be a set of gauge fields that
take values in some Lie subsuperalgebra $\k$ of the Lie
superalgebra $\s$. Then the gauged $\mathcal{N}=1$ WZNW action is
\begin{equation}\nonumber
 \begin{split}
  S[\G,A,\bar A]\ &= \ S_{\rm WZNW}^{\mathcal{N}=1}[\G]+
\frac{1}{\pi}\int d^2z d^2\theta\
  \bigl((A,\G^{-1}\bar D\G) \bigr. \\[2mm]
  & \hspace*{.5cm} \bigl. - (D\G\G^{-1},\bar A) +
(A,\bar A)- (G^{-1}A G,\bar A)\bigl)\, .
 \end{split}
\end{equation}
This action is invariant under the following gauge transformation
\begin{equation}
 \begin{split}
  \G\ &\rightarrow \ {\HH}\G\HH^{-1}\, , \\[2mm]
   A\ &\rightarrow \ \Ad(\HH)A-\HH^{-1}D\HH\, ,\\[2mm]
 \bar A\ &\rightarrow \ \Ad({\HH})\bar A-\HH^{-1}\bar D\HH\,
 \end{split}
\end{equation}
for $\HH \in K$. Thus the above action describes an $\mathcal{N}=(1,1)$
world-sheet supersymmetric $S/K$ supercoset. It is convenient to
gauge fix this symmetry such that
\begin{equation}\label{eq:supercosettrafo}
 A \ = \ D \HH \HH^{-1}\qquad,\qquad
 \bar A \ = \ \bar D \bar\HH \bar\HH^{-1}\ \ .
\end{equation}
Thereby, we can embed our coset model into the $\mathcal{N}=1$ WZNW model on
the product supergroup $S \times K$,
$$
 \int \mathcal{D}\G\mathcal{D}A\mathcal{D}\bar A\ e^{-S[\G,A,\bar A]}\ = \ \mathcal{J}
\int \mathcal{D}\G\mathcal{D}\HH\ e^{-S[\G]+S[\HH]}\,
$$
for some constant $\mathcal{J}$ as explained in
\cite{Tseytlin:1993my}. The gauge fixing procedure requires to
introduce additional ghost fields. They come in four different
kinds. There are $\dim \k^{\bar 0}$ fermionic ghosts and $\dim
\k^{\bar 1}$ bosonic ones, each contributing a central charge $c =
-2$ and $c = +2$, respectively. These all have $\mathcal{N}=1$
superpartners, i.e.\ there are $\dim \k^{\bar 0}$ bosonic ghosts
with central charge $c = -1$ and $\dim \k^{\bar 1}$ fermionic ones
with central charge $c=1$. Taking all these into account, the
ghost sector contributes  $c_{\text{ghosts}} = -3 \sdim \k$ so
that the total central charge is
\begin{eqnarray*}
 && \hspace*{-8mm} c(S/K)\ = \\[2mm]  & & \hspace*{-5mm}= \,
\left(\frac32-\frac{h^\vee_\s}{k}\right)\, \sdim \s
 + \left( \frac32 + \frac{h^\vee_\k}{k}\right) \sdim \k
 - 3\, \sdim \k \\[2mm] & & \hspace*{-5mm} =
 \ \left(\frac32-\frac{h^\vee_\s}{k}\right)\, \sdim \s
 - \left( \frac32 - \frac{h^\vee_\k}{k}\right) \sdim \k\ \ .
\end{eqnarray*}
The total Virasoro field $T_{\text{total}} = T_{\s \times \k} +
T_{\text{ghost}}$ possesses an $\mathcal{N}=1$ superpartner
$G_{\text{total}} $. Both these fields descend to the state space
of the coset model. The latter is obtained by computing the
cohomology of the BRST operator $Q$. One may show that $T_{\text{total}}$
and $G_{\text{total}}$ are in the same cohomology class
as the Virasoro element $T_{S/K}$ and its superpartner $G_{S/K}$
in the coset conformal field theory. Details on how this works in
$\mathcal{N}=1$ WZNW cosets $S/K$ of bosonic groups can be found in
\cite{Rhedin:1995um,FigueroaO'Farrill:1995pv}. The generalization
of these constructions to supergroups is entirely straightforward.
In the case of Lie groups, Kazama and Suzuki used the current
symmetry to show that some of the $\mathcal{N}=1$ WZNW cosets admit an $\mathcal{N}=2$
superconformal algebra \cite{Kazama:1988qp}. Their construction
may also be embedded into the product theory. In fact, it suffices
to show that the $\mathcal{N}=1$ superconformal algebra of the WZNW model on
$S \times K$ admits an extension to $\mathcal{N}=2$. The corresponding
fields of the $\mathcal{N}=2$ superconformal algebra receive additional
contributions from the ghost sector to form a total $\mathcal{N}=2$ algebra
whose basic $G^\pm_{\text{total}}$ and $U_{\text{total}}$ reside
in the same cohomology class as the associated fields in the coset
model. Our goal is to extend the analysis of Kazama and Suzuki to
the case in which $S$ and $K$ are Lie supergroups. According to
the remarks we have just made, all we need to do is to exhibit an
$\mathcal{N}=2$ superconformal algebra in the $\mathcal{N}=1$ WZNW model on the
product $S \times K$.

\section{Super Manin triples}

Throughout this paper, $\g$ denotes a (not necessarily simple) Lie
super-algebra with a non-degenerate supersymmetric invariant
bilinear form $(\,\cdot\, , \,\cdot\, )$. In our application to
the WZNW coset $S/K$, the Lie superalgebra $\g$ is given by $\g =
\s \oplus \k$. The form on $\g$ is determined by the form $(\,
\cdot\, , \, \cdot \,)_\s$ on $\s$ that we use to construct the
action. On $\g = \s \oplus \k$ it is given by
$$ ((X_1,Y_1),(X_2,Y_2)) \ = \ (X_1,X_2)_\s-(Y_1,Y_2)_\s $$
for all $X_i \in \s$ and $Y_i \in \k \subset \s$. As we shall show
below, possible $\mathcal{N}=2$ extensions of the $\mathcal{N}=1$ superconformal
algebra in the $S/K$ WZNW model are classified by special triples
$(\g,\aplus,\am)$. Here, $\apm$ denote two Lie subalgebras such that
\begin{equation}
    \g\ = \ \aplus\oplus\am\,\ .
\end{equation}
We call such a triple $(\g,\aplus,\am)$ a super Manin triple if
the Lie subsuperalgebras $\apm$ are isotropic, i.e.\
\begin{equation} (\apm,\apm) \ = \
0 \ \ .
\label{eq:fermionicflat} \end{equation}
For later use we also introduce the subspace $\azero$ of the Lie
superalgebra $\g$ by
\begin{equation}
 \azero \, := \, \{x\,\in\,\g\,|\,(x,y)\,=\,0\, \forall\,
 y\,\in\,[\aplus,\aplus]\cup[\am,\am]\}.
\end{equation}
Super Manin triples contain all the structure constants we shall
employ later to define the fields that generate the $\mathcal{N}=2$ super
Virasoro algebra. Before we extract the required constants,
let us discuss one series of such super Manin triples that will
become particularly important below. \medskip

\noindent {\it Example:} The most important super Manin triples we
shall exploit arise from Lie superalgebras $\g = \s$, i.e. $K = \{ e\}$ .
Let us suppose that the even part $\g^{\bar 0}$ of $\g$
splits into two bosonic subalgebras $\g^{\bar 0} =\g^{\bar
0}_a\oplus\g^{\bar 0}_b$ of equal rank. This condition applies
to the Lie superalgebras $\g=gl(n|n),psl(n|n),sl(n|n\pm1)$ and
$\g=osp(2n+1|2n),osp(2n|2n)$. In all these examples, the bilinear form of
the Cartan subalgebra of one of these subalgebras is positive
definite while the other one is negative definite (with a proper choice of real form).
Consequently, we can perform an isotropic decomposition of the
Cartan subalgebra
\begin{equation}
    \h \ = \ \h_+\oplus \h_-\, .
\end{equation}
In order to extend the decomposition of $\h$ to an isotropic
decomposition of $\g$ we recall that any Lie superalgebra admits a
triangular decomposition  into the Cartan subalgebra $\h$, the
subalgebra of the positive root spaces $\nn_+$ and the subalgebra
of negative root spaces $\nn_-$:
\begin{equation}\label{eq:triangulardecomposition}
 \g\ = \ \nn_-\oplus\h\oplus\nn_+\, .
\end{equation}
Hence the
triple $(\g,\aplus=\h_+\oplus\nn_+,\am=\h_-\oplus\nn_-)$ is a super
Manin triple, i.e.\ it satisfies the condition
\eqref{eq:fermionicflat}. We also note that the derived
subalgebras $[\apm,\apm]$ of $\apm$ are contained in $\nn_\pm$ and
consequently,
\begin{equation}\label{eq:azerocartan}
 \azero \ \supseteq \ \h\, .
\end{equation}
There exist many other super Manin triples, in particular when the
Lie superalgebra $\g$ is not simple. \medskip

Before we can turn to the $\mathcal{N}=2$ superconformal algebra we need to
extract a few structure constants that characterize the super
Manin triple. Let us pick some basis $x_i$ of the Lie superalgebra
$\aplus$. With the help of our bi-linear form $(.,.)$ we can then fix
a dual basis $x^i$ of $\am$ such that $(x_i,x^j) = \delta_{i}^j$.
Our choice of basis implies that the Lie bracket takes the
following form
\begin{equation}
    \begin{split}
        [x_i,x_j]\ &= \ {c_{ij}}^kx_k\\
        [x^i,x^j]\ &= \ {f^{ij}}_kx^k\\
        [x_i,x^j]\ &= \ {c_{ki}}^jx^k+{f^{jk}}_ix_k\ . \\
    \end{split}
\end{equation}
Here, the first two equations involve the structure constants
${c_{ij}}^k$ and ${f^{ij}}_k$ of $\aplus$ and $\am$, respectively.
The last equation follows from the first two. Let us also
introduce the projection operators $\Pi_\pm: \g \rightarrow \apm$
from the Lie superalgebra $\g$ to the two summands $\apm$.

In addition to the structure constants $c$ and $f$, our
construction of the $\mathcal{N}=2$ algebra will involve a special element
$\tilde \rho \in \g$ that is defined through
\begin{equation}\label{eq:rhotilde}
 \tilde\rho \ : \, = \ - [x^i,x_i]\ = \ (-1)^i{f^{ik}}_ix_k+(
-1)^i{c_{ki}}^ix^k\, .
\end{equation}
The Jacobi
identities for the two Lie subsuperalgebras $\apm$ as well as for the full Lie
superalgebra $\g$ imply that
\begin{equation}
        \tilde\rho \ \in \ \azero \qquad \ \text{and}\qquad\
        [\tilde\rho_+,\tilde\rho_-]\ = \ 0 \, ,
\end{equation}
where $\tilde \rho_\pm = \Pi_\pm \tilde \rho \in \apm$ is the
image of $\tilde \rho$ under the projection map $\Pi_\pm$. The
element $\tilde \rho$ determines a map $D = - \Pi_+ [\tilde
\rho,.]:\aplus \rightarrow \aplus$. When acting on the basis elements
$x_i$ it reads
\begin{equation}\label{eq:derivation}
    \begin{split}
        Dx_i \ :&= \ - \Pi_+ [\tilde\rho,x_i] \ = \ D^l_i\,  x_l
       \\[2mm]
       \mbox{where} \quad & \quad  D_i^l \ :\, = (-1)^{mn}\ {c_{mn}}^l{f^{mn}}_i\ . \\
    \end{split}
\end{equation}
The supertrace of the map $D$ is related to the length of $\tilde
\rho$ through
\begin{equation}
               \str(D) \ = \ -(\tilde\rho,\tilde\rho)\, .
\end{equation}
Any Lie superalgebra admits a canonical (often degenerate) graded
symmetric invariant bilinear Killing form. Since it also appears
in the structure constants of the current algebra, we shall
briefly evaluate the Killing form through the structure constants
$c$ and $f$. For any given choice of the basis, the Killing form
reads
\begin{equation}
    \langle X^a,X^b\rangle \ = \ -(-1)^n {C^{na}}_m{C^{mb}}_n\, .
\end{equation}
When both $X^a,X^b$ are in the same Lie subsuperalgebra $\apm$,
the Killing form on $\g$ reduces to twice the Killing form of $\apm$,
\begin{eqnarray}
\langle x_i, x_j\rangle & = & - 2(-1)^n {c_{ni}}^m {c_{mj}}^n \ = \
\kappa_{ij} \\[2mm]
\langle x^i, x^j\rangle & = & - 2(-1)^n {f^{ni}}_m {f^{mj}}_n \ = \
\kappa^{ij} \ .
\end{eqnarray}
When the two elements $X^a$ and $X^b$ are taken from different
subsuperalgebras $\apm$, the Killing form reads
\begin{equation}\label{eq:killingofg}
    \begin{split}
        \langle x_i,x^j\rangle \ &= \ {\kappa_{i}}^j
        \ = \ 2A_i^j+D_i^j\\[2mm]
         \text{where} \ \ \ \ & \ \ A_i^j \ = \
         (-1)^{mn}{c_{ni}}^m{f^{nj}}_m \ . \\
    \end{split}
\end{equation}
The matrix $D$ was defined in eq.\ \eqref{eq:derivation}. This terminates our
preparations.

\section{$\mathcal{N}=2$ superconformal algebra}

Let us begin by introducing the basic fields and their operator
product expansions. If we denote by $J_i(z)$ and $J^i(z)$ the
chiral affine currents corresponding to the generators $x_i$
and $x^i$, their operator products are \cite{Di Vecchia:1984ep}
\begin{equation}\label{eq:currentOPE}
    \begin{split}
        J_i(z)J_j(w)\ &\sim \
        \frac{\frac{1}{2}\kappa_{ij}}{(z-w)^2}+\frac{{c_{ij}}^kJ_k(w)}{(z-w)}\\[2mm]
         J_i(z)J^j(w)\ &\sim \
        \frac{{\delta_i}^j+\frac{1}{2}\kappa_i^j}{(z-w)^2}+
        \frac{{f^{jk}}_iJ_k(w)+{{c_{ki}}^j}J^k(w)}{(z-w)}\\[2mm]
        J^i(z)J^j(w)\ &\sim \
        \frac{\frac{1}{2}\kappa^{ij}}{(z-w)^2}+\frac{{f^{ij}}_kJ_k(w)}{(z-w)}
        \\[-6mm]\end{split}
\end{equation}
where $\langle x_i,x_j\rangle = \kappa_{ij}$ etc. are the entries
of the Killing form we determined at the end of the previous
section. The terms involving $\kappa$ arise because we had to
shift the metric by the Killing form in eq.\ \eqref{eq:N=1action}.
Operator product expansions of the fields $\chi^i$ and $\chi_i$
take the form
\begin{equation}\label{eq:fermionOPE}
    \begin{split}
        \chi_i(z)\, \chi_j(w)\ &\sim \ 0\\[2mm]
        \chi_i(z)\, \chi^j(w)\ &\sim \ \frac{{\delta_i}^j}{(z-w)}\\[2mm]
        \chi^i(z)\, \chi^j(w)\ &\sim \ 0\ .\\
    \end{split}
\end{equation}
All these fields have conformal weight $h(\chi_i) = 1/2 = h(\chi^i)$.
The pair $\chi_i$ and $\chi^i$ form a bosonic $\beta \gamma$ system with $c =
-1$ when $|i| = 1$ and they generate a fermionic $bc$ system of
central charge $c =1$ when $|i|=0$.

\newcommand{\nol}{:\!}
\newcommand{\nor}{\!\!:}
Let $(\g,\aplus,\am)$ be a super Manin triple of a Lie superalgebra
$\g$ such that the condition \eqref{eq:fermionicflat} holds. We
now want to build a U(1) current $U$, the Virasoro field $T$ and
two fermionic currents $G^\pm$ of weight $h = 3/2$ such that they
obey the algebra of an $\mathcal{N}=2$ superconformal symmetry. We begin
with the current $U$,
\begin{equation}
    U(z) = \ \nol \chi^i\chi_i \nor+ \tilde \rho^k J_k +
    \tilde \rho_k J^k +D^i_j\nol \chi^j\chi_i\nor\, .
\end{equation}
Here, we have extracted the numbers $\tilde \rho_i$ and $\tilde
\rho^i$ from our element $\tilde \rho \in \g$ through
\begin{eqnarray*}
\tilde \rho_k & := & (\tilde \rho, x_k) \ = \
(-1)^i {c_{ki}}^i\\[2mm]
\tilde \rho^k & := & (\tilde \rho, x^k) \ = \
(-1)^{i} {f^{ik}}_i
\ .
\end{eqnarray*}
The Virasoro tensor $T$ takes the usual form
\begin{equation}
    T(z)\ = \ \frac{1}{2}(\nol J^iJ_i\nor + (-1)^i \nol J_i J^i\nor +
    \nol \del \chi^i\chi_i\nor -\nol \chi^i\del \chi_i\nor )\,
\end{equation}
as a sum of the Sugawara tensor of the affine superalgebra at
level $k+h^\vee$ and the Virasoro tensor of the free fields $\chi_i$ and $\chi^i$.
Finally, we introduce the two super-currents by \footnote{We thank
Yasuaki Hikida for discussions that led us to the particular sign combinations in the second
term that are needed for applications to type II superalgebras. See also the recent paper \cite{Giribet:2009eb}.}
\begin{equation}
    \begin{split}
        G^+(z)\ &= \, J_i\chi^i-\frac{1}{2}(-1)^{i+ij}{c_{ij}}^k\,\nol
        \chi^i\chi^j\chi_k\nor\\[2mm]
        G^-(z)\ &= \, J^i\chi_i-\frac{1}{2}(-1)^{j+ij}{f^{ij}}_k\,
        \nol \chi_i\chi_j\chi^k\nor \ . \\
    \end{split}
\end{equation}
We claim that $(U,T,G^\pm)$ form an $\mathcal{N}=2$ superconformal algebra
of central charge
\begin{equation}
 c \ = \ \frac{3}{2}\, \sdim\,\g+3\, \str D\, .
\end{equation}
For simple Lie supergroups $\g$, $\str D = - h^\vee \sdim\g/3k$ so
that the value of the central charge agrees with what we had
spelled out in section 2. The fields $T,G^\pm$ and $U$ extend the
$\mathcal{N}=1$ superconformal symmetry of the $S \times K$ WZNW model. In
fact, the Virasoro field $T = T_{\s \times \k}$ and its $\mathcal{N}=1$
superpartner $G = G^+ + G^- = G_{\s \times \k}$ agree with the
$\mathcal{N}=1$ superconformal structure of the WZNW on the product $S\times
K$. As we explained at the end of section 2, all fields must be
augmented by the standard contributions from the ghost sector
before they descend to the desired $\mathcal{N}=2$ superconformal algebra of
the coset model.

In order to prove the claim that the four currents $T, U$ and
$G^\pm$ form an $\mathcal{N}=2$ superconformal algebra one has to compute
their operator products. This has been done carefully in
\cite{Creutzig:2009zz}. After inserting the operator products
\eqref{eq:currentOPE} and \eqref{eq:fermionOPE} of the constituent
fields $J(z)$  and $\chi(z)$, the resulting expressions can be
simplified with the help of the Jacobi identity, as in the case of
bosonic groups $G$.

For the key example of a super Manin triple that we described in
the previous section, $\str D = 0$ and hence the central charge of
the associated $\mathcal{N}=2$ superconformal algebra is  given by $c =
\frac{3}{2}\sdim\,\g$. Some of the supercosets that admit a
super Manin triple are listed in a table below, along with the
central charge.
\begin{table}[h]\label{table:supercosets}
\begin{center}
 \begin{tabular}{ | c | c | c |  }
    \hline
    $S$ & $K$ $\vphantom{\Bigl(\Bigr)}$ & $c\bigl(S/K\bigr)$ \\[1mm] \hline\hline
    GL($n|n$)  & GL($n-m|n-m$)$\vphantom{\Bigl(\Bigr)}$  & 0\\  \hline
    GL($n|n$)  & SL($n-m|n-m\pm1$)$\vphantom{\Bigl(\Bigr)}$  & 0\\  \hline
    PSL($n|n$)  & PSL($n-m|n-m$)$\vphantom{\Bigl(\Bigr)}$  & 0 \\  \hline
    PSL($n|n$)  & SL($n-m|n-m\pm1$)$\vphantom{\Bigl(\Bigr)}$ & -3 \\  \hline
    SL($\tilde n|n$)$\ \, \tilde n > n$ &
    SL($\tilde n-m|n-m$)$\vphantom{\Bigl(\Bigr)}$  & 0\\  \hline
\end{tabular}\caption{{\em Incomplete list of $\mathcal{N}=2$ superconformal
supercosets $S/K$ with central charge $c\bigl(S/K\bigr)$. In all
cases we assume that $n > m \geq 0$.}}
\end{center}
\end{table}

There exist more $\mathcal{N}=2$ superconformal algebras, which are obtained
from the previous ones through a deformation by an element $\alpha$ in
$\azero$. Consider an element $\alpha=p^ix_i+q_ix^i \in
\azero$ where $p^i,q_i$ are Grassmann elements of grade $|i|$. It follows from the very definition of $\azero$ that the
components $p^i$ and $q_i$ must satisfy
\begin{equation}\label{eq:alpha}
    {c_{ij}}^kq_k \ = \ {f^{ij}}_kp^k \ = \ 0\, .
\end{equation}
We employ the element $\alpha$ to deform the fields of the $\mathcal{N}=2$
superconformal algebra as follows
\begin{equation}
    \begin{split}
        U_\alpha(z) \ &= \ U(z)+p^i\, I_i(z)-(-1)^iq_i\, I^i(z)\\[2mm]
        T_\alpha(z) \ &= \ T(z)+\frac{1}{2}(p^i\, \del I_i(z)+
        (-1)^iq_i\, \del I^i(z))\\
    \end{split}
\end{equation}
where we used the following set of level $k$ Lie superalgebra currents
\begin{equation*}
    \begin{split}
        I_i \, &= \, J_i-(-1)^{i+ij}{c_{ij}}^{k} \nol \chi^j\chi_k\nor -\frac{1}{2}(-1)^{ik}{f^{jk}}_i \nol \chi_j\chi_k\nor\\[2mm]
        I^i \, &= \, J^i-(-1)^{j+ij}{f^{ij}}_k \nol \chi_j\chi^k\nor -\frac{1}{2}(-1)^{ij}{c_{jk}}^{i}\nol \chi^j\chi^k\nor\, . \\
    \end{split}
\end{equation*}
The expressions for the deformed supercurrents $G^\pm$ are a bit
simpler
\begin{equation}
    \begin{split}
        G^+_\alpha \ &= \ G^++q_i\, \del \chi^i \\[2mm]
        G^-_\alpha \ &= \ G^-+p^i\, \del \chi_i\, . \\
    \end{split}
\end{equation}
Since we want $G^\pm$ to remain fermionic under the deformation,
we required $\alpha$ to be bosonic. The central charge of the
deformed algebra is
$$ c_\alpha = c-6(-1)^iq_ip^i\, .$$
The deformed $\mathcal{N}=2$ structure extends a deformation of the original
$\mathcal{N}=1$ superconformal algebra. It is relevant in particular for the
discussion of models that are obtained from the WZNW model by
Hamiltonian reduction.

\section{The $\mathcal{N}=1$ WZNW models on GL(1$|$1)}

In the following section we would like to illustrate our
constructions in the simplest model, the $\mathcal{N}=1$ WZNW model on the
supergroup GL(1$|$1). The GL(1$|$1) WZNW model has been discussed in
\cite{Rozansky:1992rx, Rozansky:1992td, Schomerus:2005bf, Creutzig:2007jy,
Creutzig:2008ek, Creutzig:2008an}. The Lie superalgebra \gl\ is generated by
elements $E,N, \psi_\pm$ such that
$$ [N, \psi_\pm] \ =\ \pm \psi_\pm \quad ,
\quad [\psi_+,\psi_-] \ = \ E\ \ $$
and $E$ commutes with all other generators. It comes equipped with
an invariant bilinear form $(.,.)$ whose non-vanishing entries are
$$ (E,N) \ = \ k \quad , \quad (\psi_+,\psi_-) \ = \ k\ \ . $$
Written in terms of the various component fields, the action of
the $\mathcal{N}=1$ GL(1$|$1) WZNW model is
\begin{equation} \label{glact}
 \begin{split}
 S \ = \ &\frac{1}{2\pi}\int d^2z\ \bigl(
k\del X\bar\del Y+k\del Y\bar\del X +\del Y\bar\del Y+\\[1mm]
&\qquad+2e^Y\del\cp\bar\del\cm+\chi^N\bar\del\chi^E+\chi^E\bar\del\chi^N+\\[2mm]
&\qquad+\chi^+\bar\del\chi^--\chi^-\bar\del\chi^++\bar\chi^N\del\bar\chi^E+\\[2mm]
&\qquad+\bar\chi^E\del\bar\chi^N+\bar\chi^+\del\bar\chi^--\bar\chi^-\del\bar\chi^+\bigr)\, .
\end{split}
\end{equation}
Note the additional term $\del Y\bar\del Y$  which is not present
in the usual $\mathcal{N}=0$ WZNW model on GL(1$|$1). This term is due to the
shift of the bi-linear form by the Killing form (see our comment in
section 2). The Lie supergroup GL(1$|$1) is not simple but solvable
and its superalgebra has a degenerate but non-zero Killing form
with the only non-vanishing entry being
$$ \langle N,N \rangle \ = \ 2 \ .
$$
The model \eqref{glact} has a gl(1$|$1) current algebra symmetry
generated by four currents $J^E,J^N,J^\pm$. Their $\mathcal{N}=1$
superpartners will be denoted by $\chi^E,\chi^N,\chi^\pm$. We note
that the Cartan algebra of \gl\ has two generators $E$ and $N$
which are isotropic. Hence, we can introduce a super Manin triple
$($\gl$,\aplus,\am)$ through
\begin{equation}
\aplus \ :\, = \ \mbox{\it span\/}(E,\psi_+) \quad , \quad \am  \
:\,=\ \mbox{\it span\/} (N,\psi_-) \ .
\end{equation}
It follows that the subspace $\azero$ is spanned by $E,N$ and
$\psi_-$. We shall work with the basis $x_1 = E/\sqrt{k}, x_2 =
\psi_+/\sqrt{k}$ and
$x^1 = N/\sqrt{k}, x^2 = \psi_-/\sqrt{k}$ such that the only non-vanishing structure
constants are
$$ {f^{12}}_2 \ =\  -\frac{1}{\sqrt{k}} \ = \  -{f^{21}}_2\ .$$
Consequently, the element $\tilde \rho$ takes the form $\tilde
\rho = -E/k$ and hence $D =0$. According to our general formulas,
the U(1)-current $U$ and the two super-currents $G^\pm$ are given
by
\begin{equation}
\begin{split}
      U \ & = \ \chi^N\chi^E+\chi^-\chi^+-\frac{J^E}{\sqrt{k}}\\[2mm]
       G^+\ &=\ J^E\chi^N+J^+\chi^- \\[2mm]
    G^-\ &=\ J^N\chi^E+J^-\chi^+ -\frac{1}{\sqrt{k}}\chi^E\chi^+\chi^-\, .
\end{split}
\end{equation}
One can construct another anti-holomorphic $\mathcal{N}=2$ superconformal
algebra out of the anti-holomorphic currents, exactly in the same
way as we did in the holomorphic case.

As we have briefly reviewed in the introduction, the $\mathcal{N}=2$
superconformal algebra determines two topological conformal field
theories that are obtained through $A-$ and $B-$twist. The
physical states of the $B$-twisted model form the so-called
$(c,c)$ ring while those of the $A$-twisted model are in the
$(c,a)$ ring. We would like to determine these two state spaces
for the example at hand. Let us recall that any representative
$\phi$ of a $(c,c)$ or $(c,a)$ state must obey
\begin{equation} \label{eq:CRcond}
2\Delta(\phi)+\epsilon u(\phi) \ = \ 2\bar\Delta(\phi)+
\bar\epsilon' \bar u(\phi) \ = \ 0\,
\end{equation}
 where
$\Delta(\phi), \bar\Delta(\phi), u(\phi)$ and $\bar u(\phi)$ are
the conformal dimensions and U(1)-charges of the field $\phi$.
States in the $(c,c)$ ring correspond to $\epsilon = 1 =
\epsilon'$ while those in the $(c,a)$ ring are associated with
$\epsilon = 1 = - \epsilon'$.

All representatives of the $(c,c)$ and $(c,a)$ ring are based on
the components of the fields
\begin{equation}\label{vpen}
 \Phi_{n+1} \ = \
   \left(\begin{matrix} e^{inY}
    & i\cm e^{inY} \\ i\cp e^{inY} &
    \cm \cp e^{inY}
    \end{matrix}\right) \ \ \ \mbox{ for } \ \ n \in \
    \mathbb{R}\ .
\end{equation}
These correspond to harmonic functions on the supergroup
GL(1$|$1), i.e.\ to functions that are annihilated by (some power
of) the Laplacian. Only the first column is in the kernel of $Q_B
= G^+_0$ and $\bar Q_B = \bar G^+_0$. The complete $(c,c)$ ring is
then spanned by products of the form
$$ \left( e^{inY} , i\cp e^{inY} \right) \times
    \left(1,  \chi^N,\bar \chi^N,\chi^N
       \bar \chi^N\right)\, .$$
Let us note that operators involving the bosonic fields $\chi^-$
and $\bar \chi^-$ contribute to the kernel of $Q_B$ and $\bar
Q_B$, but not to the cohomology since they are exact. For the
$(c,a)$ ring, a similar analysis can be performed. In this case,
the kernel of $Q_A = G^+_0$ and $\bar Q_A = \bar G^-_0$ in the space of
atypical fields \eqref{vpen} contains the constant function only.
The $(c,a)$ ring is then represented by the following four fields
$$ \left(1, \chi^N, \bar \chi^E,\chi^N\bar\chi^E\right)\, . $$
It is not difficult to verify (see e.g.\ \cite{Schomerus:2005bf})
that neither the $(c,c)$ nor the $(c,a)$ ring depend on the level
$k$. We also note that many states satisfying eqs.\
\eqref{eq:CRcond} are not part of the chiral ring of the model.
This is in sharp contrast to the situation in unitary models
\cite{Lerche:1989uy}.

\section{Conclusions and Open Problems}

In this work we exhibited $\mathcal{N}=2$ superconformal symmetries for a
large class of $\mathcal{N}=1$ WZNW models. Our constructions generalize
previous studies of bosonic models \cite{Spindel:1988sr,Kazama:1988qp}
to the case of target superspaces. One of the main new features is the
existence of $\mathcal{N}=2$ superconformal symmetry in $\mathcal{N}=1$ WZNW models
of simple supergroups such as PSL(N$|$N) or OSP(2N+1$|$2N).
As a concrete example, we analyzed the $\mathcal{N}=1$ WZNW model on GL(1$|$1)
and computed its (anti-)chiral ring. The contributions to the
(anti-)chiral ring were all associated with states in atypical
representations of the target space supersymmetry. This feature
is expected to extend to higher supergroup target spaces.

The case of PSL(N$|$N) is particularly interesting. Since
PSL(N$|$N) possess vanishing dual Coxeter number, the
corresponding WZNW model can be deformed away from the WZ point
while preserving conformal symmetry
\cite{Bershadsky:1999hk,Quella:2007sg}. In other words, the WZNW
models on PSL(N$|$N) are special points in a one-parameter family
of conformal field theories with unbroken global symmetry. The
same holds for the $\mathcal{N}=1$ version of these models. Given that those
deformed models still possess chiral Virasoro fields, one may
wonder about the fate of the $\mathcal{N}=2$ superconformal symmetry. We
believe that the fields $G_\pm$ and $U$ also remain chiral under
the deformation. The issue will be addressed in forthcoming work.

Among the coset theories with non-trivial denominator, the
superspace generalization of $\mathcal{N}=2$ minimal models are of
particular interest. The compact and non-compact versions are
given by the two cosets PSL(1,1$|$2)/SL(1$|$2) and
PSL(1,1$|$2)/SL(1,1$|$1). Both theories possess central charge
$c=-3$, regardless of their level.

There are a number of other extensions of the present
work that deserve a closer investigation. One of them is to
incorporate world-sheets with boundary. The $\mathcal{N}=1$ WZNW models on the
supergroups PSL(N$|$N), GL(N$|$N) and SL(N-1$|$N), for example,
are all known to possess two families of maximally symmetric
boundary conditions \cite{Creutzig:2008ag}. In
\cite{Creutzig:2009zz}, one of them was shown to descend to the
A-twisted model while the other is consistent with the B-twist.
Cosets with non-trivial denominator possess a richer structure.
Finally, one might also wonder whether some of the $\mathcal{N}=(2,2)$
theories we discussed here allow for $\mathcal{N}=(4,4)$ superconformal
symmetry. The answer turns out to be positive. We shall describe
the exact conditions and consequences in a forthcoming paper.

\begin{acknowledgements}
We thank Nicolas Behr, Nathan Berkovits, Constantin Candu, Manfred Herbst,
Kentaro Hori, Vladimir Mitev, David Ridout, Hubert Saleur
and in particular Yasuaki Hikida for conversations and comments.
\end{acknowledgements}

\end{document}